\documentclass[conference]{IEEEtran}
\IEEEoverridecommandlockouts
\usepackage{cite}
\usepackage{amsmath,amssymb,amsfonts}
\usepackage{algorithmic}

\usepackage{amsmath} 

\usepackage{algorithm}
\usepackage{hyperref}
\usepackage{booktabs}
\usepackage{subcaption}

\usepackage{graphicx}
\usepackage{textcomp}
\usepackage{xcolor}
\def\BibTeX{{\rm B\kern-.05em{\sc i\kern-.025em b}\kern-.08em
    T\kern-.1667em\lower.7ex\hbox{E}\kern-.125emX}}
\begin{document}

\title{Parallel Instance Filtering for Malware Detection\\
\thanks{This work was supported by the OP VVV MEYS funded project CZ.02.1.01/0.0/0.0/16\_019/0000765 ''Research Center for Informatics'' and by the Grant Agency of the CTU in Prague, grant No. SGS21/142/OHK3/2T/18 funded by the MEYS of the Czech Republic.}}

\author{\IEEEauthorblockN{Martin Jure\v{c}ek and Olha Jure\v{c}kov\'{a}}
\IEEEauthorblockA{\textit{Department of Information Security} \\
\textit{Faculty of Information Technology, Czech Technical University in Prague, Czechia}\\
\{martin.jurecek,jurecolh\}@fit.cvut.cz}
}

\maketitle

\begin{abstract}
Machine learning algorithms are widely used in the area of malware detection. With the growth of sample amounts, training of classification algorithms becomes more and more expensive. In addition, training data sets may contain redundant or noisy instances. The problem to be solved is how to select representative instances from large training data sets without reducing the accuracy. This work presents a new parallel instance selection algorithm called Parallel Instance Filtering (PIF). The main idea of the algorithm is to split the data set into non-overlapping subsets of instances covering the whole data set and apply a filtering process for each subset. Each subset consists of instances that have the same nearest enemy. As a result, the PIF algorithm is fast since subsets are processed independently of each other using parallel computation. 
We compare the PIF algorithm with several state-of-the-art instance selection algorithms on a large data set of 500,000 malicious and benign samples. The feature set was extracted using static analysis, and it includes metadata from the portable executable file format. Our experimental results demonstrate that the proposed instance selection algorithm reduces the size of a training data set significantly with the only slightly decreased accuracy. The PIF algorithm outperforms existing instance selection methods used in the experiments in terms of the ratio between average classification accuracy and storage percentage. 
\end{abstract}


\section{Introduction} \label{sec:introduction}
Machine learning (ML) algorithms are widely employed in the area of malware detection \cite{ucci2019survey}. AV-Test Institute \cite{avtest2022avtest} registers over 450 000 new malware samples and potentially unwanted applications every day. This indicates that antivirus companies must deal with a large amount of training data. Training data sets usually contain many irrelevant samples, which do not contribute to improving detection accuracy. In addition, some samples are manually labeled by malware analysts, which may result in incorrect labels. 

The training phase of ML algorithms is time expensive. Reducing training time and cleaning training sets are two main reasons to incorporate instance selection algorithms \cite{olvera2010review} into malware detection systems. The goal of an instance selection algorithms is to remove redundant and noise instances from a given training data set while keeping or even increasing the classification accuracy. 

Feature selection and extraction algorithms are among the most important preprocessing steps in data science tasks. Instance selection is becoming increasingly relevant due to the large number of samples that are constantly being generated. Since the computational complexity of the most instance selection algorithms is at least $O(n^2)$ where $n$ is a number of instances \cite{garcia2010democratic}, most of the algorithms are not applicable for larger data sets.

In this paper, we propose a parallel instance selection algorithm based on splitting the whole training data set into disjoint subsets and applying a filtering rule to each subset. The union of the reduced subsets then forms the reduced training data set. We focus on the binary classification problem, where we try to distinguish between malicious and benign programs. The experiments were conducted for the data from the portable executable (PE) file format  \cite{microsoftPE}, the most widely used file format for malware samples for personal computers.

The main contributions are as follows.  We propose a novel parallel instance selection algorithm called  Parallel Instance Filtering (PIF). The algorithm is easy to implement and fast since it processes independent subsets of the data set rather than the whole data set.  We compared the PIF algorithm with four state-of-the-art instance selection algorithms for medium-sized training data sets and large-sized training data sets containing up to 400,000 high-dimensional samples. While the computational times of state-of-the-art methods were too high for processing more than 100,000 samples, the PIF algorithm could process any subset of our data set. We used the stratification strategy for the large training data sets to compare the PIF algorithm with other instance selection algorithms. For both medium and large data sets, the PIF algorithm achieved the results of the state-of-the-art algorithms considered in our experiments.

This paper is organized as follows. Section \ref{background_related_work} provides a brief introduction to malware detection and reviews some instance selection algorithms. Section \ref{proposed} presents our proposed instance selection algorithm. Experimental setup is given in Section \ref{setup}, and Section \ref{evaluation} describes the experimental results. Finally, Section \ref{conclusion} concludes the paper and presents the future work.

\section{Background and Related Work} \label{background_related_work}

This section provides a brief background for malware detection and also presents the state-of-the-art instance selection algorithms used in the experiments. The rest of this section reviews some instance selection techniques suitable for large data sets.

\subsection{Background}

\subsubsection{Malware Detection}

Malware authors adapt their malicious programs frequently to bypass antivirus programs that are regularly updated. Packing, encryption, polymorphism, metamorphism, and other code obfuscation techniques \cite{monnappa2018learning} are widely used by malware authors to evade signature detection techniques \cite{kephart1994automatic}.  

Packing and encryption are two of the most popular techniques. These techniques apply compression or encryption algorithms to avoid signature-based detection and static analysis \cite{nath2014static}. Polymorphic and metamorphic malware can change their code with each new generation. When such malware is executed, it can be obfuscated again to avoid signature-based detection.

Obfuscation can be applied on several layers, such as code, a sequence of instructions, or binary content. Even a small malware change can invalidate an existing signature and cause the modified malware not to be detected immediately after the modification. As a result, training data sets may contain many samples that are similar to each other with respect to a similarity function defined on a feature set. However, while the number of training samples increases, the learning speed decreases. In addition, similar training samples do not provide much new information for machine learning classifiers. Therefore, it is convenient to reduce the training data set by selecting only representative samples.

Malware detection techniques can be classified into two categories depending on how code is analyzed: static and dynamic analysis. The static analysis aims at searching for information about the structure and data in the file. The disassembly technique is an example of a static analysis technique used to extract various features from the executables. The dynamic analysis \cite{or2019dynamic} aims to examine a program executed in a real or virtual environment. 

Many malware researchers have focused on ML algorithms to defeat obfuscation techniques and detect unknown malware. Malware detection problems are often defined as classification or clustering problems. The choice of ML algorithms depends on whether training data is labeled or not and depends on data representation. The works \cite{raff2020survey} and \cite{ucci2019survey} survey the most popular machine learning techniques for malware classification in the Windows operating system.

Our work is based on static analysis. The most important drawback is that data captured from the static analysis does not describe the complete behavior of a program since the program is not executed. Since dynamic analysis involves running the program, information from this kind of analysis is more relevant than information from static analysis. However, dynamic analysis is more time-consuming than static analysis, and several anti-virtual machine technologies \cite{afianian2019malware} can evade detection systems based on dynamic analysis. Since dynamic analysis could be impractical for a large volume of samples that come to antivirus vendors daily, the static analysis still has its place in malware detection systems. A comparison of these two approaches is presented in \cite{damodaran2017comparison}.

\subsubsection{Instance Selection Algorithms}

The goal of instance selection algorithms is to reduce training data sets by selecting only representative instances while keeping (and possibly improving) accuracy. The work \cite{olvera2010review} reviews several instance selection algorithms and discusses their properties. In \cite{garcia2012prototype}, the authors provide extensive comparative experiments of instance selection algorithms evaluated using the $k$-nearest neighbor (KNN) classifier. The work \cite{brighton2002advances} compares instance selection algorithms for 30 data sets from different domains, and the authors report that neither instance selection algorithm consistently outperforms the other.

The main disadvantage of instance selection algorithms is high computational complexity. Thus, they are not applicable to large-size data sets. To avoid this drawback, the stratified strategy \cite{cano2005stratification} is combined with instance selection algorithms and achieves good results compared to instance selection algorithms without stratification.

The stratified strategy splits the training set randomly into disjoint subsets of equal size, maintaining class distribution within each subset. Then, an instance selection algorithm is applied to each subset, and the resulting reduced training set is obtained by joining the reduced subsets. Stratification allows us to run any instance selection algorithm on small subsets instead of a large set of instances. As a result, reducing of small subsets can be run in parallel. We use the stratified strategy in our experiments when we compare the PIF algorithm with other instance selection algorithms for large-sized data sets.

There are several ways to categorize instance selection algorithms. We divide instance selection algorithms, as in \cite{olvera2010review}, into two groups: filter methods and wrapper methods. While the selection criterion of wrapper methods is based on the accuracy of the classifier, the selection criterion of filter methods is not based on a classifier but rather on the values of the feature vector. 

In this work, we focus on wrapper methods. In the rest of this section, we briefly describe the methods we used in the experiments. The detailed description of these methods can be found in \cite{olvera2010review}. 

Condensed nearest neighbor (CNN) \cite{cover1967nearest} is probably the first published instance selection algorithm. The objective of CNN is to find a \textit{consistent set}, i.e., a subset $S$ of training set $T$ that correctly classifies (using $S$ as the training set) every instance in $T$ with the same accuracy as when $T$ is used as the training set. CNN does not produce a minimal \textit{consistent set} if redundant instances were added to the consistent set in the early steps. CNN is order-dependent and sensitive to random initialization of \textit{consistent set}. 

Edited Nearest Neighbor (ENN) \cite{wilson1972asymptotic} is focused on filtering noisy instances in a training set. An instance is removed if its class does not coincide with the majority class of its $k$ nearest neighbors. We use ENN as a noise filter in our proposed instance selection algorithm described in the next section.
 
Wilson and Martinez introduced the Decremental Reduction Optimization Procedure (DROP) family methods \cite{wilson1997instance} which are very popular as they outperform several previously proposed state-of-the-art methods. DROP methods are based on the concept called \textit{associate}. An instance $p$ is the \textit{associate} of the instance $q$ if $q$ is one of $k$ nearest neighbors of the instance $p$. DROPs methods remove an instance $q$ if its associates can be classified without $q$. Among five methods, DROP1-DROP5, we selected DROP3 since it achieved the best mix of storage reduction and generalization accuracy \cite{wilson2000reduction}.

Iterative Case FilteringAlgorithm (ICF) \cite{brighton1999consistency} uses two special sets, \textit{Coverage} and \textit{Reachable}, that are related to the concepts of \textit{neighborhood} and \textit{set of asociates} used in DROP algorithm. The definitions are as follows:
$$\textrm{Reachable(x)} = \{y \in S: y \in \textrm{LocalSet($x$)}\}$$
$$\textrm{Coverage(x)} = \{y \in S: x \in \textrm{LocalSet($y$)}\}$$

where the LocalSet($x$) is defined by the set
$$\{y \in S: \textrm{cl(y)} = \textrm{cl(x)} \land \mathcal{D}(x,y)<\mathcal{D}(x,NE_x)\},$$

where $cl(x)$ denotes the class of $x,$ $\mathcal{D}$ denotes the distance function, and $NE_x$ denotes the nearest enemy of $x$, i.e., a nearest instance to $x$ having different class label. When the cardinality of the corresponding reachable set is greater than the cardinality of the corresponding coverage set, then the instance is removed from the training set.  The first step of ICF consists of the noise filtering scheme based on ENN. The reason is such that if some instance is noisy, then the cardinality of both a reachable set and a coverage set will be one.

Modified Selective Subset (MSS) \cite{barandela2005decision} is based on concept of \textit{selective subset} which is defined as follows:

A subset $S$ of the training set $T$ is a selective subset, if
\begin{enumerate}
	\item $S$ is consistent
	\item all instances in $T$ are closer to any instance of $S$ of the same class than to any enemy in $T$
\end{enumerate}

The authors of MSS proposed an iterative procedure to find the modified selective subset, which is defined as the subset of the training set $T$ which contains, for every instance $x$ in $T$, the element of its neighborhood that is the closest to the nearest enemy of $x$.

\subsection{Related Work}

In malware detection, there is a distinct lack of experimentation with instance selection methods applied on large and real-world data sets from the Windows environment. To the best of our knowledge, this paper is one of the first works on instance selection for malware detection. Therefore, this section reviews only instance selection algorithms that were applied to data from different areas, such as biology or medicine. 

MapReduce framework \cite{dean2008mapreduce} is another way how to accelerate instance selection algorithms. The framework was used in work \cite{triguero2015mrpr} for instance selection with data sets up to 5.7 millions of instances. Locality-sensitive hashing (LSH) \cite{indyk1998approximate} can be used to support  the calculation of nearest neighbors \cite{andoni2006near}. The work \cite{arnaiz2016instance} proposed an instance selection algorithm with linear complexity using LSH to find similarities between instances. 

In \cite{garcia2010democratic}, the authors proposed a method for scaling up instance selection algorithms. The original data set is partitioned into disjoint subsets using the theory of Grand Tour \cite{asimov1985grand}. An instance selection algorithm is applied to each subset, and the results are combined using a voting scheme to achieve a low testing error and high storage reduction.

\section{Proposed Method} \label{proposed}

This section presents a novel instance selection method called the Parallel Instance Filtering (PIF) algorithm. The PIF algorithm is a wrapper method and belongs among decremental methods, i.e., instances are removed from the original data set according to some selection rule.

The PIF algorithm consists of three components:
\begin{enumerate}
	\item Wilson Editing noise-filtering procedure - incorrectly classified instances using KNN classifier are removed.
	\item partitioning the data set - creating disjoint subsets $S_{NE}$ corresponding to nearest enemies $NE$,
	\item applying filtering rule - filter out instances from each subset $S_{NE}$ according to a filtering rule.
\end{enumerate}

In the first step of the PIF algorithm, we apply the Wilson Editing noise-filtering procedure, which is presented in Algorithm \ref{noisefiltering}. 

\begin{algorithm}
\caption{Wilson Editing noise-filtering procedure}
\label{noisefiltering}
\begin{algorithmic}[1]
\REQUIRE $T$ - original training set, $k$ - number of nearest neighbors
\ENSURE filtered training set $S$
\FORALL{points $x$ from $T$}
\STATE find $k$ nearest neighbors of $x$ in $T$ (excluding $x$)
\STATE classify $x$ using KNN
\ENDFOR
\FORALL{points $x$ from $T$}
\IF{$x$ is classified incorrectly} 
\STATE remove $x$ from $T$
\ENDIF
\ENDFOR
\RETURN $T$
\end{algorithmic}
\end{algorithm}

This procedure is similar to ENN \cite{wilson1972asymptotic}, and its strategy is to filter out incorrectly classified instances. This includes noisy instances as well as boundary instances. If the procedure was not included in the first step of the PIF algorithm, then this would result in increasing the number of subsets and decreasing their size.

Note that the Wilson Editing noise-filtering procedure for filtering noise and smoothing borders is commonly used in the first step of instance selection algorithms. This procedure is, for example, included in the DROP3 and the ICF algorithms.

However, Wilson Editing noise-filtering procedure cannot remove redundant internal instances which do not provide any information for the classifier in the training process. Therefore, the PIF algorithm applies a filtering rule to remove such instances that are irrelevant for training the classifier.

In the second step of the PIF algorithm, all instances in the training data set that were not filtered out are then divided into disjoint subsets according to their nearest enemies. If some instance has two or more distinct nearest enemies, we can randomly choose one of them and omit the others.

The reason why we always can decompose the training data set $T$ into disjoint subsets covering the whole data set is as follows. Let $x,y$ be two instances from $T$ and consider the following relation $\sim_{NE}$ on $T$:
$$y_1 \sim_{NE} y_2 \textrm{ if and only if } NE_{y_1}=NE_{y_2}$$

If each instance $y$ has a unique nearest enemy $NE_y$ then it is easy to see that the relation $\sim_{NE}$ is an equivalence relation. Therefore, equivalence classes $[x]_{NE}=\{y\in T \mid x=NE_y \}$ form a decomposition of $T$. 

After decomposition of the training data set each subset is reduced independently by applying the following filtering rule: \\
 
\begin{figure}
\centering
\includegraphics[scale=0.7]{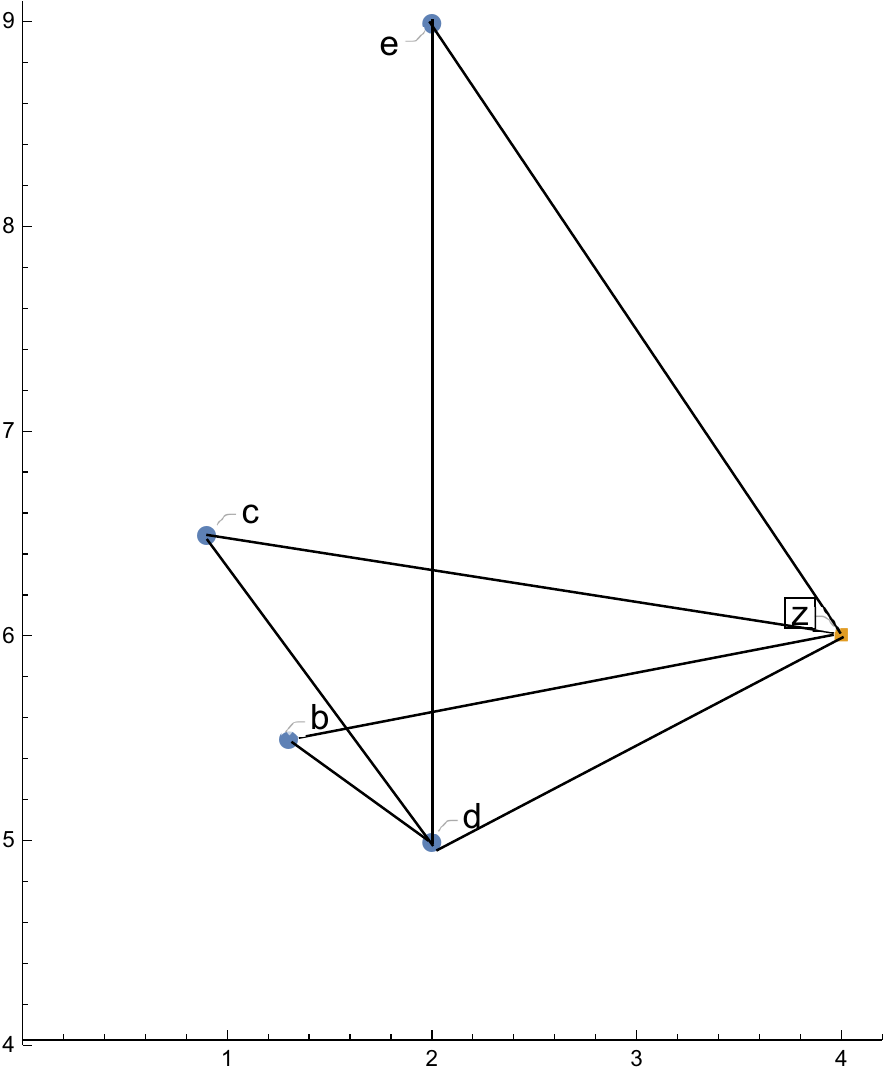}
\caption{Demonstration of the filtering rule used in the PIF algorithm.}
\label{fig:1}
\end{figure}

\textit{remove an instance $y$ from the subset $S_{NE}$ if and only if there is $x \in S_{NE}$ such that}
\begin{equation} \label{filt_rule}
\mathcal{D}(y,NE) \geq  \max \{\mathcal{D}(y,x),\mathcal{D}(x,NE)\}
\end{equation}

where $\mathcal{D}$ is a distance function defined for the feature vectors from the training set $T.$ This filtering rule is shown in Fig. \ref{fig:1} where the point $z$ highlighted in a square is the nearest enemy for the remaining points. The figure demonstrates that the points $b$ and $c$ will be removed, however, the points $d$ and $e$ are retained. Since the Wilson Editing noise-filtering procedure did not remove the points $d$ and $e$, they are probably not noisy instances or outliers. For example, the reason for retaining the point $e$ is that there is no point $x$ such that:
\begin{itemize}
	\item $x$ is closer to $z$ than the point $e$, and
	\item $e$ is closer to $x$ than to its nearest enemy $z$.
\end{itemize}

The PIF algorithm is described in Algorithm \ref{PIF}. The algorithm contains the parameter $m$, which determines the smallest size of the subsets to which the filtering rule (\ref{filt_rule}) is applied.

\begin{algorithm}
\caption{Parallel Instance Filtering Algorithm}
\label{PIF}
\begin{algorithmic}[1]
\REQUIRE $T$ - original training set , $m$ - minimal size of a subset 
\ENSURE reduced training set
\STATE perform Wilson Editing noise-filtering procedure for $T$
\FORALL{points $x$ from $T$}
\STATE compute nearest enemy $NE_x$ 
\STATE compute a subset $S_{NE_x}$
\ENDFOR
\FORALL{nearest enemies $NE$} 
\IF{$|S_{NE}| \geq m$}
\FORALL{points $y$ from $S_{NE}$}
\STATE find $x \neq y$ in $S_{NE}$ such that \\ $\mathcal{D}(y,NE) \geq  \max \{\mathcal{D}(y,x),\mathcal{D}(x,NE)\}$
\IF{such $x$ was found}
\STATE remove $y$ from $T$
\ENDIF
\ENDFOR
\ENDIF
\ENDFOR
\RETURN $T$
\end{algorithmic}
\end{algorithm}

From the implementation point of view, while we compute all-to-all distances and find $k$ nearest neighbors for each instance $x$, we also save all nearest enemies, and for each enemy, $NE_x$, create subset $S_{NE_x}$ of instances having $NE_x$ as its nearest enemy. 

 The PIF algorithm runs in parallel since each of the following parts of the algorithm can be implemented in a parallel way:
 \begin{itemize}
 \item finding $k$ nearest neighbors, and all nearest enemies $NE$ and as a result, computing subsets $S_{NE}$ of instances corresponding to $NE$
 \item applying KNN classifier used in the Wilson Editing noise-filtering procedure 
 \item filtering out instances for each subset $S_{NE}$ independently according to the rule (\ref{filt_rule}).
 \end{itemize}
 
\begin{figure*}
\begin{subfigure}[b]{0.31\linewidth}
\centering
\includegraphics[ width=\linewidth]{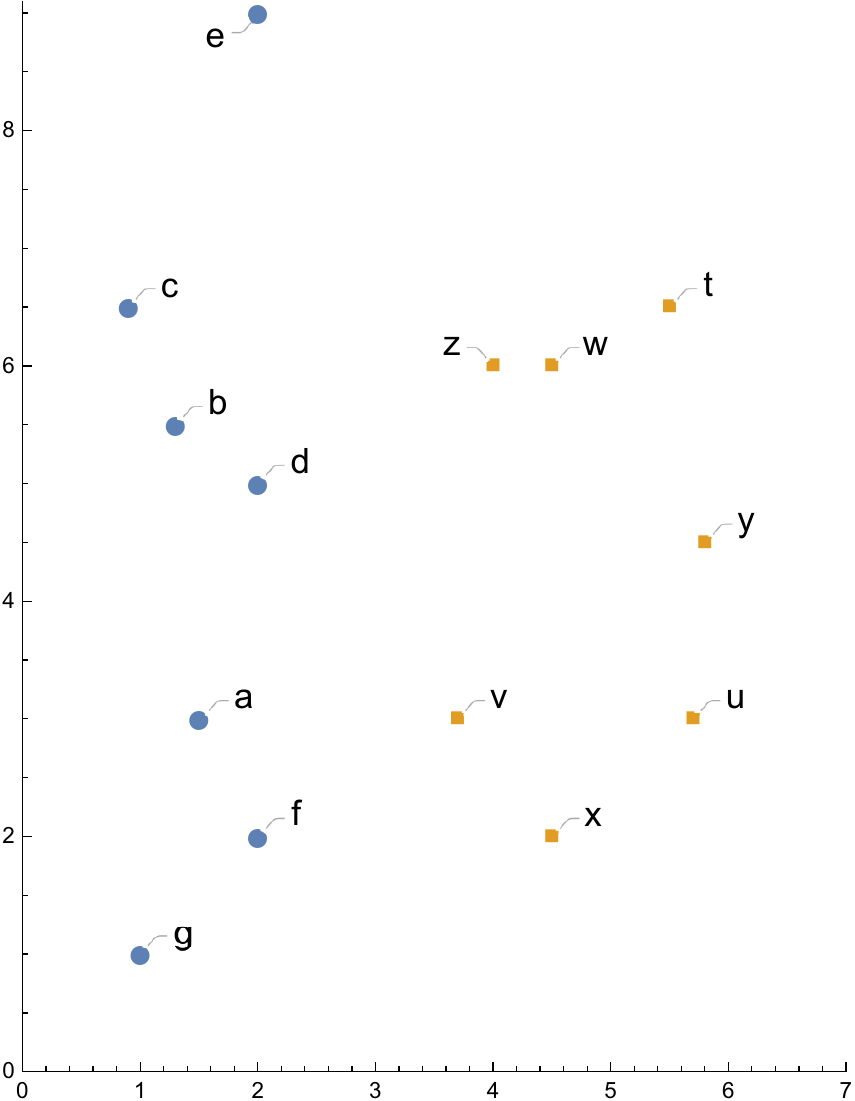}
\caption{Original data set.}
\end{subfigure}
\hfill
\begin{subfigure}[b]{0.31\linewidth}
\centering
\includegraphics[width=\linewidth]{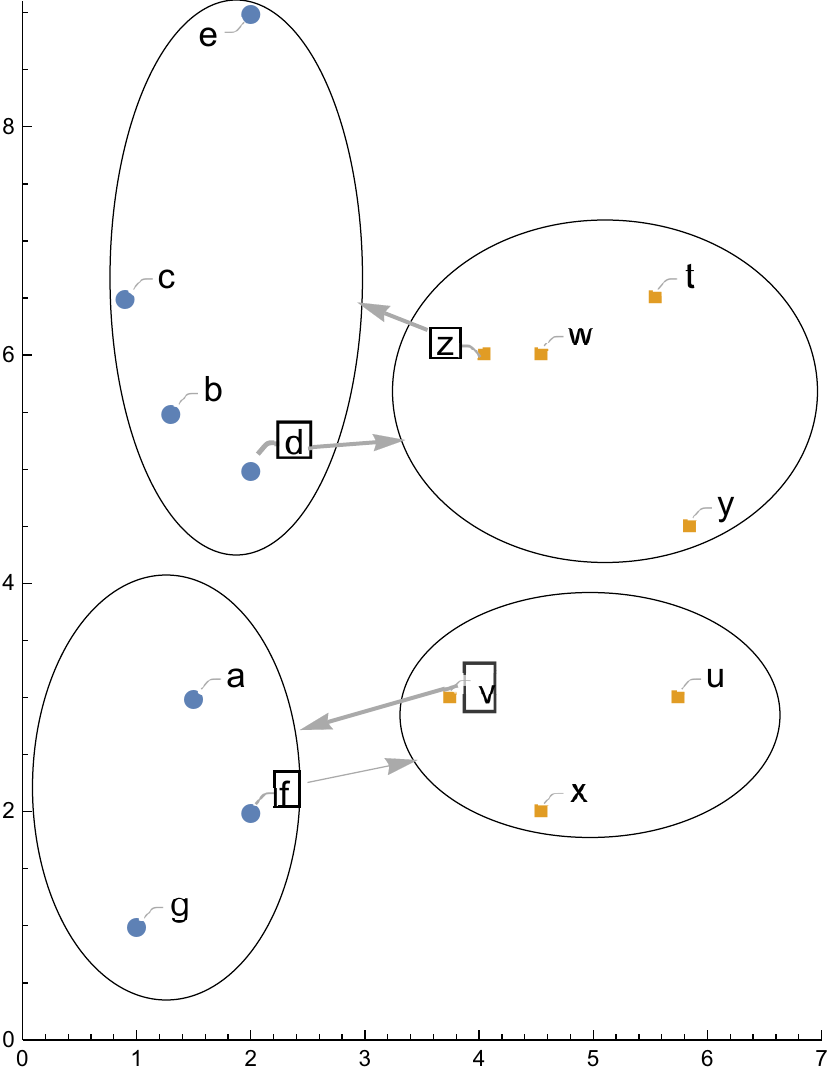}
\caption{Disjoint subsets.}
\end{subfigure}
\hfill
\begin{subfigure}[b]{0.31\linewidth}
\centering
\includegraphics[width=\linewidth]{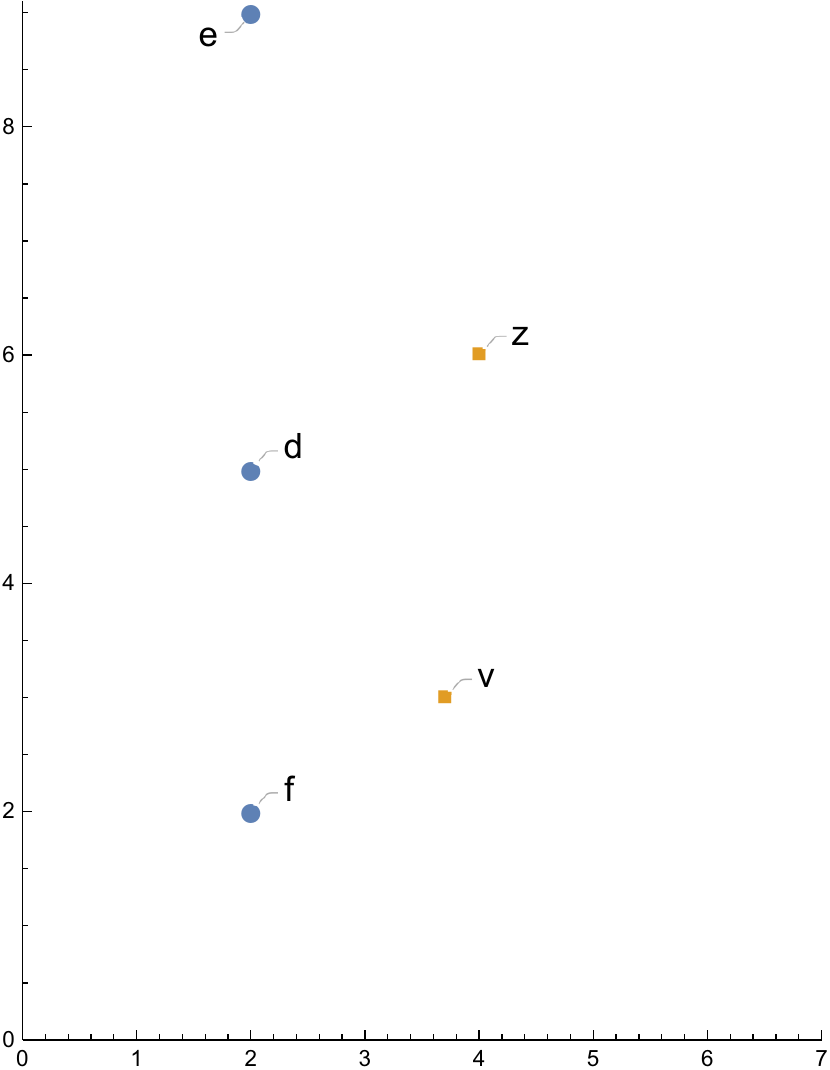}
\caption{Reduced data set.}
\end{subfigure}
\caption{Demonstration of the PIF algorithm for a training data set consisting of four disjoint subsets. The nearest enemies are highlighted in squares.} \label{fig:2}
\end{figure*}

Fig. \ref{fig:2} demonstrates the PIF algorithm for the data set that consists of the four disjoint subsets.

\section{Experimental Setup} \label{setup}

The data set used in our experiments contains real-world data from 500,000 Windows programs in the PE file format, where half are malicious and half are benign programs. The malicious and benign programs were obtained from the industrial partner’s laboratory, and the open-access Virusshare repository \cite{virusshare}.

This work is based on static analysis, which aims to search 
for information about the structure and data in the file without actually running the program. The features used in our work are metadata from the PE file format \cite{microsoftPE}, which is the most widely used file format for malware samples for personal computers. A PE file consists of headers and sections that encapsulate the information necessary to manage the executable code. To classify an executable file in the PE format, we extract static format information and translate it into a feature vector suitable for classification. To extract features from the PE file format, we used Python module \verb|pefile| \cite{carrera2017pefile}. This module extracts all PE file features into an object from which they can be easily accessed. In total, we extracted 370 features based on static information only, i.e., without running the program. Nominal features were transformed into numeric values, and all numeric values were normalized using \textit{min-max normalization}. We then select the 75 most relevant features for distinguishing between malicious and benign files using the Recursive Feature Elimination for Logistic Regression algorithm, where both methods are implemented in the Scikit-learn library \cite{scikitlearn}. A detailed description of the preprocessing of features can be found in our previous work \cite{jurevcek2021application}.

In all experiments, the parameter $m$ of the PIF algorithm was set to 2. As a result, the filtering rule (\ref{filt_rule}) was applied to any possible subsets.

The performance of instance selection algorithms on the test set is measured using the following standard methods. The most intuitive and commonly used evaluation metric in machine learning is accuracy (ACC). It is defined on a given test set as the percentage of correctly classified instances. We evaluated instance selection algorithms using the KNN classifier. The authors of the DROP family methods recommended to use a small odd number of nearest neighbors, $k$, such as 1, 3, or 5. In our experiments, we set the $k=3$.

The second measure is a storage percentage (R). It shows what percent of the original (non-reduced) training set was retained by the instance selection algorithm. 

Some instance selection algorithms outperform others in terms of storage percentage, while some instance selection algorithms outperform others in terms of the accuracy of the KNN classifier applied to the reduced data set. Therefore, we also consider the ratio between average accuracy and storage percentage. Our experiments presented in the following section demonstrate that such a ratio was the highest for the PIF algorithm applied to large data sets.

\section{Experimental Results} \label{evaluation}

In this section, we present several experiments where we compared the PIF algorithm with existing instance selection algorithms for different data sizes. All experiments in this work were executed on a single computer platform having two processors (Intel Xeon Gold 6136, 3.0GHz, 12 cores each), with 64 GB of RAM running the Ubuntu server 18.04 LTS operating system. 

All experiments presented in this paper were repeated 20 times for randomly chosen training and test data sets from a total set of 500,000 instances. We kept the ratio between the size of the training set and the size of the test set at 80:20. The results in all tables and figures show the average values of storage percentage and accuracy.

Eight different data set sizes were considered for the evaluation. We divided them into two categories:

\begin{itemize}
	\item medium data sets: 5K, 10K, 20K, 50K, and 100K of samples,
	\item large data sets: 200K, 300K, and 400K of samples,
\end{itemize}

where the capital letter K represents times one thousand (e.g., 100K = 100,000). 

\subsection{Evaluation on medium data sets}

A summary of our experiment for medium size data sets is provided in Table~\ref{tab:1}. The KNN algorithm that retains 100 \% of the training set is also included for comparison. Table~\ref{tab:1} demonstrates that when the size of data sets decreases, then classification accuracy also decreases. 

\begin{table*}
\renewcommand{\arraystretch}{1.3}
\caption{Accuracy and storage percentage represented in \%. The results presented in bold represent the highest accuracy and the lowest storage percentage for the data sets.} \label{tab:1}
\begin{footnotesize}
\begin{center}
\begin{tabular}{c|c|c|c|c|c|c|c|c|c|c|c|c|} 
\cline{2-13}
& \multicolumn{2}{|c|}{KNN} & \multicolumn{2}{|c|}{CNN}& \multicolumn{2}{|c|}{DROP3}& \multicolumn{2}{|c|}{MSS} & \multicolumn{2}{|c|}{ICF} & \multicolumn{2}{|c|}{PIF}\\
\hline
\multicolumn{1}{|c|}{size} & ACC & R & ACC & R & ACC & R & ACC & R & ACC & R & ACC & R \\
\hline
\multicolumn{1}{|c|}{5K} & 92.72 & 100 & 90.64 & 17.94 & 88.64 & \textbf{13.70} & 91.28 & 18.9 & 84.72 & 21.7 & \textbf{91.52} & 16.2\\
\multicolumn{1}{|c|}{10K}  & 93.04 & 100 & 91.80 & 15.21 & 90.16 & \textbf{10.90} & \textbf{92.72} & 16.1 & 90.04 & 17.7 & 91.84 & 13.39\\
\multicolumn{1}{|c|}{20K} & 94.50 & 100 & 93.12 & 12.73 & 92.46 & \textbf{8.44} & \textbf{93.20} & 13.7 & 91.10 & 17.7 & \textbf{93.20} & 11.62\\
\multicolumn{1}{|c|}{50K} & 95.98 & 100 & \textbf{95.29} & 10.24 & 93.22 & \textbf{6.74} & 94.65 & 11.2 & 92.69 & 18.3 & 94.25 & 8.74\\
\multicolumn{1}{|c|}{100K} & 96.46 & 100 & 95.51 & 8.75 & 94.71 & \textbf{5.69} & \textbf{95.88} & 9.85 & 92.77 & 17.4 & 95.63 & 7.81 \\\midrule
\end{tabular}
\end{center}
\end{footnotesize}
\end{table*}

DROP 3 achieved the lowest storage percentage for all data sets sizes, while the PIF algorithm achieved the second-lowest storage percentage. However, in terms of accuracy, DROP3 achieved the second-worst classification results. The PIF algorithm achieved the highest average accuracy for data sets with 5,000 and 20,000 samples, while it achieved the second-highest accuracy for the 10K and the 100K data sets. 

\subsection{Evaluation on large data sets}

This section summarizes the results for large data sets. The largest data set for which we evaluated all instance selection algorithms used in this work has 100,000 samples (i.e., the 100K data set). The DROP3 reached the highest computing time on this data set, exceeding 24 hours. Since the computation time of the state-of-the-art instance selection methods used in this work was too high for large data sets, we used a stratified strategy. 

Since the PIF algorithm can process individual subsets in parallel, it is suitable for processing also large sets. However, the experiments also contain the results of the PIF algorithm for the data set to which the stratification was applied. A number of subsets into which the data set is divided is given as a parameter to the stratification algorithm. We considered the following values for the parameter: 200, 400, 600, 800, 1000, 1200, and 1400. For the value 1000, we achieved the best ratio between the classification accuracy and storage percentage. As a result, we divide each data set of size $n$ into $n/1000$ disjoint subsets. The results using stratification are shown in Table \ref{tab:2}.

\begin{table*}[htbp]
\renewcommand{\arraystretch}{1.3}
\caption{Accuracy and storage percentage represented in \%. The results presented in bold represent the highest accuracy and the lowest storage percentage for the data sets. The postfix \_s in the name of the instance selection algorithm means that the stratification strategy was used.} \label{tab:2}
\begin{center}
\begin{tabular}{c|c|c|c|c|c|c|c|c|c|c|c|c|c|c|} 
\cline{2-15}
& \multicolumn{2}{|c|}{KNN} & \multicolumn{2}{|c|}{CNN\_s}& \multicolumn{2}{|c|}{DROP3\_s}& \multicolumn{2}{|c|}{MSS\_s} & \multicolumn{2}{|c|}{ICF\_s} & \multicolumn{2}{|c|}{PIF\_s}& \multicolumn{2}{|c|}{PIF}\\
\hline
\multicolumn{1}{|c|}{size} & ACC & R & ACC & R & ACC & R & ACC & R & ACC & R & ACC & R & ACC & R\\
\hline
\multicolumn{1}{|c|}{100K} & 96.46 & 100  & 93.70  & \textbf{7.67} & 93.76 & 13.91 & \textbf{95.62} & 21.71 & 93.93 & 23.60 & 94.40 & 15.87 & 95.58 & 7.81 \\
\multicolumn{1}{|c|}{200K} & 96.80 & 100 & 94.61 & \textbf{6.68} & 94.53 & 12.12 & 95.95 & 18.41 & 94.67 & 22.61 & 95.10 & 13.82 & \textbf{96.00} & 6.73  \\
\multicolumn{1}{|c|}{300K} & 97.12 & 100 & 95.32 & \textbf{6.04} & 95.25 & 10.96 & 96.20 & 16.92 & 95.11 & 22.21 & 95.64 & 12.73  & \textbf{96.52} & 6.11 \\
\multicolumn{1}{|c|}{400K} & 97.23 & 100 & 95.28 & \textbf{5.67} & 95.43 & 10.11 & 96.32 & 15.88 & 95.17 & 19.32 & 95.76 & 12.10  & \textbf{96.61} & 5.73 \\
\midrule
\end{tabular}

\end{center}
\end{table*}

The PIF algorithm achieved the highest average accuracy for all sizes except 100K and the second-lowest storage percentage for all data sets. While CNN using a stratified strategy achieved the lowest storage percentage, it also achieved the lowest accuracy for the 100K data set, the second-lowest accuracy for the 200K and 400K data set, and the third-lowest accuracy for the 300K data set.

For a data set with 100K samples, we can compare the accuracy and storage percentages of instance selection algorithms without stratification (see Tab.~\ref{tab:1}) and when stratification was used (see Tab.~\ref{tab:2}). All instance selection algorithms, except the ICF, achieved higher accuracy when stratification was not used. Regarding storage percentage, only DROP3 and MSS achieved lower storage percentage when stratification was not used.

\begin{figure}
\includegraphics[scale=0.6]{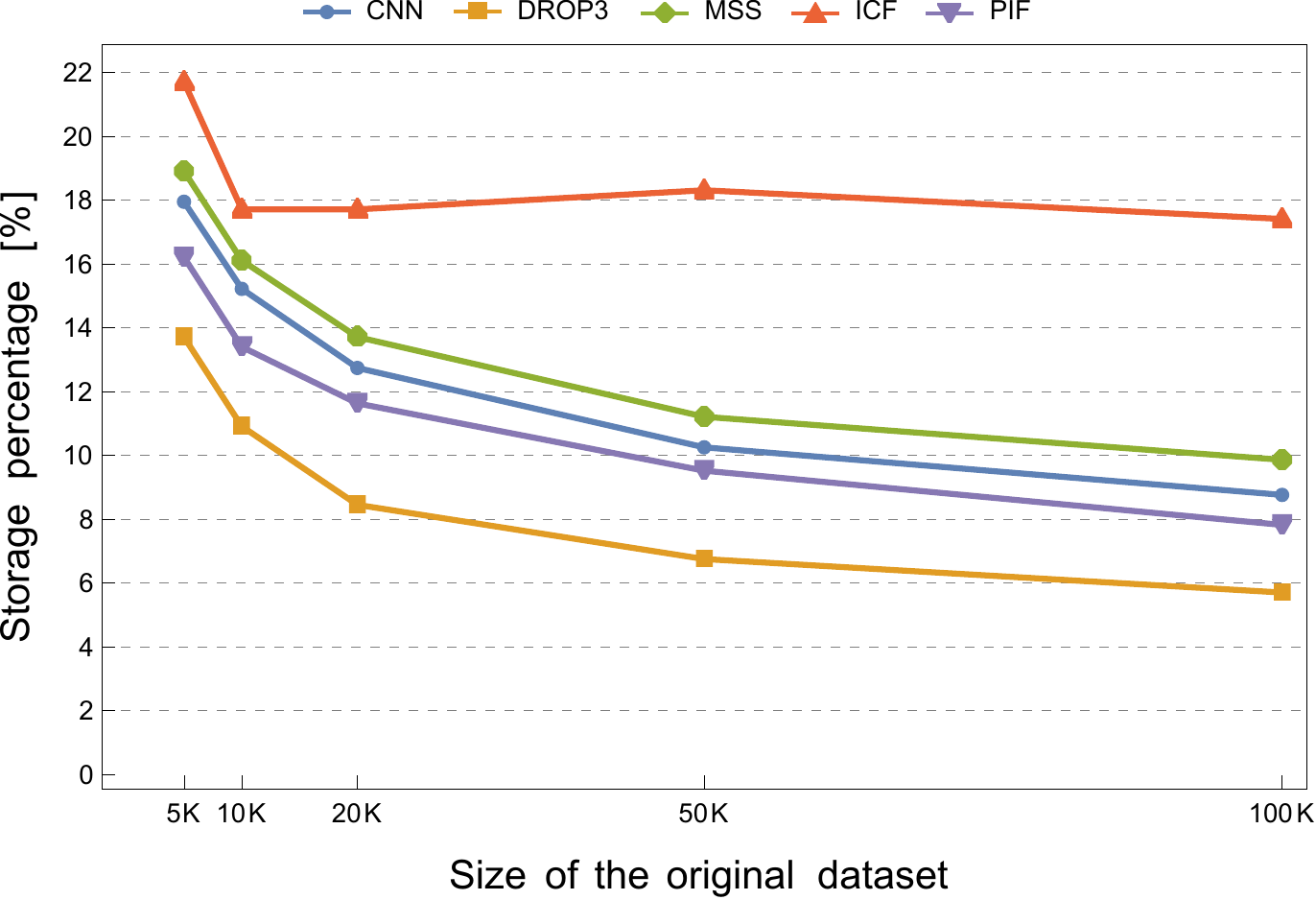}
\caption{Comparison of storage percentage of the PIF algorithm with the selected state-of-the-art instance selection algorithms.}
\label{fig:3}
\end{figure}

\begin{figure}
\includegraphics[scale=0.6]{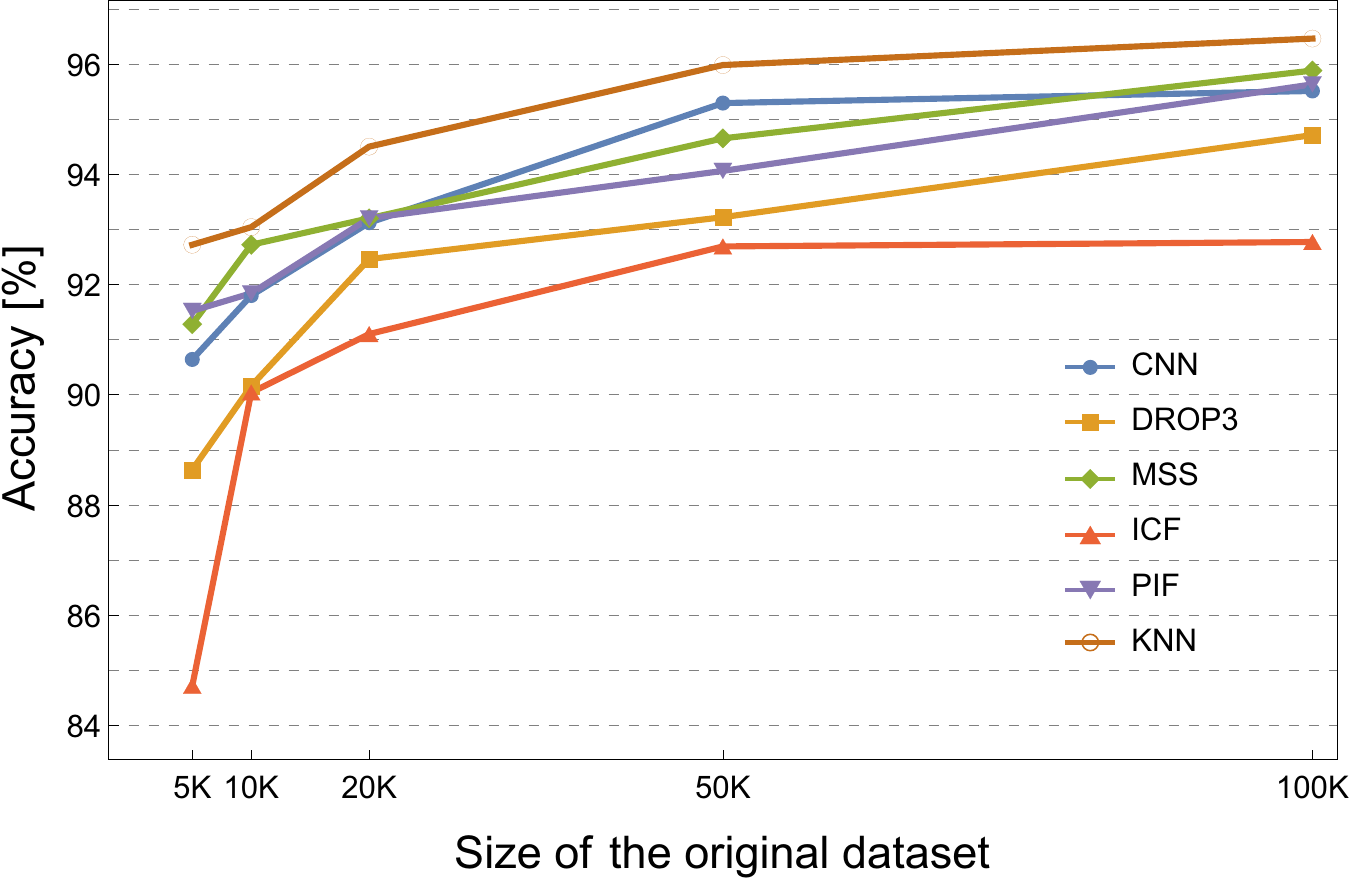}
\caption{Comparison of accuracy percentage of the PIF algorithm with the selected state-of-the-art instance selection algorithms. The KNN classifier was trained using the original (non-reduced) training data set, and all the remaining algorithms were trained using reduced training sets.}
\label{fig:4}
\end{figure}

Figures \ref{fig:3} and \ref{fig:4} illustrate the performance of the PIF algorithm for medium-sized data sets in terms of the storage percentage and the accuracy, respectively. The figures compare the results with the performance of the instance selection algorithms used in our experiments.

\begin{figure}
\centering
\includegraphics[scale=0.48]{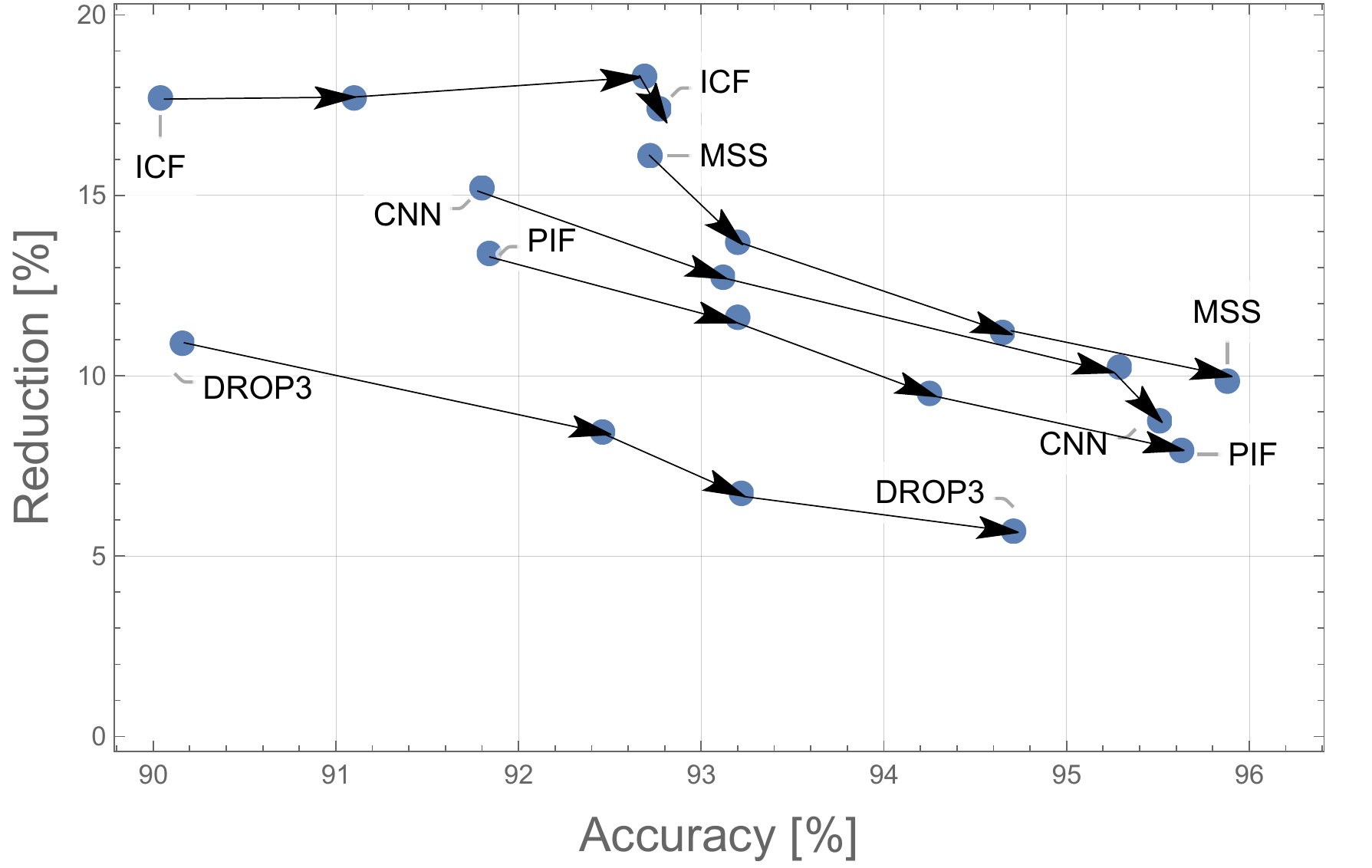}
\caption{Relation between the reduction (i.e., storage percentage) and the accuracy percentage. The results were obtained using the data set of sizes 10K, 20K, 50K, and 100K of samples.}
\label{fig:storacc1}
\end{figure}

\begin{figure}
\centering
\includegraphics[scale=0.48]{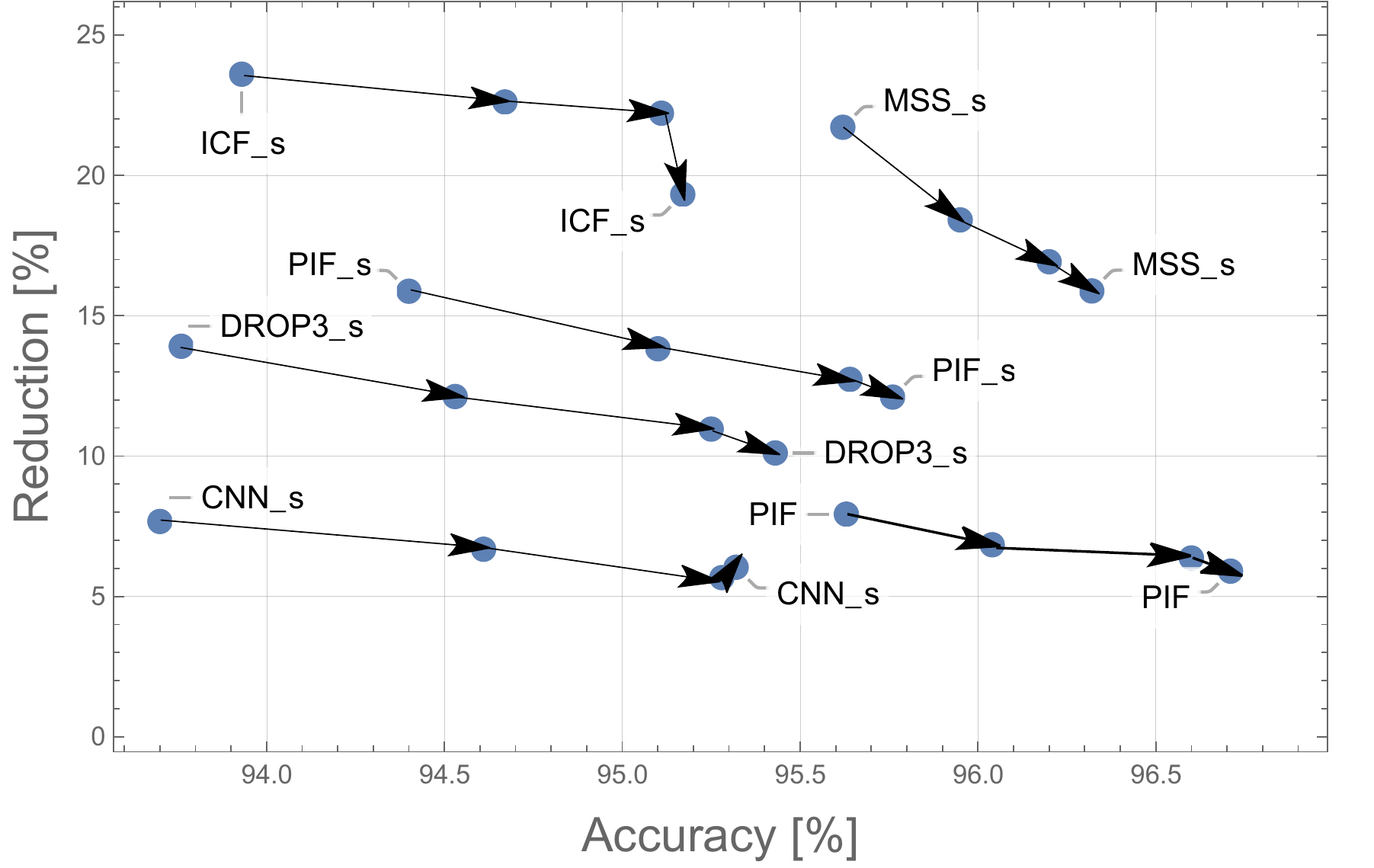}
\caption{Relation between the reduction (i.e., storage percentage) and the accuracy percentage. The results were obtained for the data set of sizes: 100K, 200K, 300K, and 400K of samples.}
\label{fig:storacc2}
\end{figure}

Figures \ref{fig:storacc1} and \ref{fig:storacc2} demonstrate the relation between the storage percentage and the accuracy for medium-sized data sets and for large-sized data sets respectively. Fig.~\ref{fig:storacc1} represents a five instance selection algorithm as an oriented polyline connecting four points that correspond (from left to right) to the following data set's sizes: 10K, 20K, 50K, and 100K. Fig.~\ref{fig:storacc2} represents the five instance selection algorithm using stratification and the PIF algorithm without stratification. Similarly, the instance selection algorithms are represented as an oriented polyline connecting four points that correspond (from left to right) to the following data set's sizes: 100K, 200K, 300K, and 400K. The figures demonstrate that the PIF algorithm achieved the results of the state-of-the-art instance selection algorithm. In addition, the PIF algorithm achieved the highest ratio between average accuracy and storage percentage for large-sized data sets.

Our implementation of the PIF algorithm reduced the 400K data set in 2 hours and 31 minutes on average, while the computational times of all other instance selection algorithms considered in this work exceeded 24 hours. One of the reasons for such a significant difference among computational times is that the PIF algorithm runs fully in parallel. In contrast, the other instance selection algorithms run in parallel only partly (typically, computing one-to-one distances among instances runs in parallel).

\section{Conclusion} \label{conclusion}

Antivirus vendors have very large data sets intended for training machine learning algorithms. As a result, the training process becomes more and more expensive. Instance selection algorithms deal with this problem by removing redundant and noise instances, and they were applied to data sets from different domains. In this paper, we examined the performance of selected instance selection algorithms for the problem of malware detection. We proposed a novel instance selection algorithm called Parallel Instance Filtering (PIF) that splits the training data set into disjoint subsets according to the nearest enemies and applies a filtering rule to each subset independently. Our algorithm is easy to implement, runs in parallel, and can significantly reduce training sets with only a small reduction in accuracy.

Several experiments were conducted on the medium and large data sets containing metadata from the portable executable file format. For a large training data set consisting of 400,000 samples, the PIF algorithm achieved a storage percentage of 5.90\%, and the accuracy using the non-reduced data set, 97.23\%, was decreased to 96.71\% using the reduced data set. Experimental results for large data sets showed that PIF outperformed the state-of-the-art methods considered in this work in terms of the ratio between the average accuracy and storage percentage. 

In future work, we plan to experiment with different filter rules for the PIF algorithm and also perform extended benchmark analysis for data sets from different domains.

\bibliographystyle{IEEEtran}
\bibliography{IEEEabrv,paper}

\end{document}